\documentclass[final,jgsp]{geom}
\usepackage{amsmath}
\usepackage{empheq}
\usepackage{amssymb}
\usepackage{comment}
\usepackage{hyperref}

\DeclareMathOperator{\Wnabla}{\overset{\rm w}{\nabla}}
\DeclareMathOperator{\Christo}{\Gamma}
\DeclareMathOperator{\WChrist}{\overset{\rm w}{\Gamma}}

\begin{document}
\head{66}{2023}

\title
{A Weyl geometric approach to  the gradient-flow equations in information geometry $^1$}
\author[Tatsuaki Wada]
{Tatsuaki Wada}
\date{}
\maketitle

\stepcounter{footnote}

\begin{abstract}[Communicated by Vladimir I. Pulov]
The gradient-flow equations with respect to the potential functions in information geometry are reconsidered from the perspective of Weyl integrable geometry.
The  pre-geodesic equations associated with the gradient-flow equations are regarded as the general pre-geodesic
equations in the Weyl integrable geometry.
\\[0.2cm]
\textsl{MSC}:  53E99,  
53Z05   
\\
\textsl{Keywords}: Dually-flat structure, gradient-flow equations, information geometry, Weyl integrable geometry
\end{abstract}

\section{Introduction}
\label{Intro}
Information geometry (IG) \cite{Amari} is a useful method exploring the fields of information science by means of differential geometry.
IG is invented from the studies on invariant properties of a manifold of probability distributions, and a modern treatment is based on
the affine differential geometry \cite{M10}.
IG has been applied to the different fields including statistical physics \cite{BR03,BH09}, statistics, dynamical systems \cite{N94,FA95} and so on.
It is known that the gradient-flow equations are useful for some optimization problems. A recent attentions is the gradient-flows in  metric spaces \cite{AGS08}.
The gradient flows on a Riemann manifold follow the direction of gradient descent (or ascent) in the landscape of a potential functional, with respect to the curved structure of the underlying metric space. The information geometric studies on the gradient systems were originally studied by Nakamura \cite{N94}, Fujiwara and Amari \cite{FA95}.
A remarkable feature of their works is that a certain kind of gradient flow on a dually flat
space can be expressed as a Hamilton flow.  Specifically, the linear differential equations 
\begin{align}
  \frac{d \eta^{\rm gf}_j }{ dt}  = -\eta^{\rm gf}_j,  \qquad j=1,2, \dots, 2 m
  \label{orig-gradEq}
 \end{align}
 coincide with 
Hamiltonian's equations for their proposed Hamiltonian 
\begin{align}
H^{\rm gf} = - \sum_{k=1}^m Q^k  P_k
\end{align}
where  $Q^k =  \eta^{\rm gf}_{2k}$ denotes the generalized position and $P_k = -(1/ \eta^{\rm gf}_{2k-1})$ the generalize momentum, respectively.
Here, each $k$-th term $- Q^k P_k$, ($k=1, \ldots, m$) in the Hamiltonian $H^{\rm gf} $ is a first integral, i.e., a conserved quantity with respect to time evolution. Since the system described by Hamilton's equations with respect to this $H^{\rm gf} $ is in an  $m$-dimensional subspace of  the $2m$-dimensional symplectic space and the system has $m$-conserved quantities, this is a completely integrable system. 
A completely integrable systems admit alternative Hamiltonian descriptions as shown by Marmo, Sparano, and Vilasi  \cite{MSV13}.
The several researches on this issue have been done from the different perspectives.
Malag\'o and Pistone \cite{MP15} studied the natural gradient flow in the mixture geometry of a discrete exponential family.
Boumuki and Noda \cite{BN16} studied the relationship between Hamiltonian flow and gradient flow from the perspective of symplectic geometries. 
The same issue was studied in \cite{WSM21,WSM22} from the perspective of geometric optics.  A path of the gradient-flow in IG can be regarded as the light path (or ray) described by the anisotropic Huygens equation.
The gradient-flows in IG are also related to the thermodynamic processes, and it is shown \cite{WSM21} that the evolutional parameter in the gradient flow equations in IG is related to the temperature of the simple thermodynamic systems based on the Hamilton-Jacobi dynamics.
The gradient-flows are related to the co-geodesic flows of the geodesic Hamiltonians \cite{WSM22}.
Furthermore the analytical mechanical properties concerning the gradient-flow equations in IG are studied in \cite{CW22}, and 
discussed  the deformations of the gradient-flow equations which lead to Randers-Finsler metrics. 
Through these studies we realize the importance of treating space and time on equal footing, which is an essence of Einstein's relativity. In order to conveniently describe physical equations in general relativity, the selection of coordinates system in curved space-time is important and non-trivial. In a suitable coordinate system, the physical equations have simple forms and clear physical meanings \cite{G18}.  We consider that the same as true for the gradient-flow equations in IG.

In this work, we reconsider the gradient-flow equations in IG from the perspective of Weyl integrable geometry.
To my knowledge this is the first result which relates the gradient-flow in IG and Weyl integrable geometry.
The rest of paper is organized as follows. Section 2 briefly review the basic of IG and the gradient-flow equations.
In Section 3, the Weyl integrable geometry is reviewed. Wely's non-metricity conditions, the pre-geodesic equations, and Weyl's gauge transformations are explained.
Section 4 is the main section, and the gradient-flow equations is related to the pre-geodesic equations in the Weyl integrable geometry.
Final section is devoted to Conclusions.
Hereafter we use Einstein's summation convention for repeated indices.

\section{Information Geometry and Gradient-Flow Equations}
\label{IG}
Here, the basic of IG and the gradient-flow equations are reviewed.

\subsection{Information Geometry}
IG \cite{Amari} is invented from the studies on invariant properties of a manifold of probability distributions.
The \textit{dually-flat structures} are important
and a statistical manifold $(\mathcal M, g, \nabla, \nabla^\star)$ is characterized
by a (psuedo-) Riemannian metric $g$, and torsion-less dual affine connections
$\nabla$ and $\nabla^\star$. For a given convex function $\Psi(\theta)$ together with its dual convex function $\Psi^\star(\eta)$,
one can construct a dually-flat structures as follows.
From the dual convex functions $\Psi^\star (\eta)$ and $\Psi (\theta)$,
the associated dual affine coordinates $\theta^i$ and $\eta_i$ 
are obtained as
\begin{align} 
\theta^i = 
\frac{\partial \Psi^\star(\eta)}{\partial \eta_i}, \qquad 
\eta_i = 
\frac{\partial \Psi(\theta)}{\partial \theta^i}
\label{theta-eta} 
\end{align}
respectively. The convex functions are Legendre dual to each to other.
\begin{align}
    \Psi^\star(\eta) = \theta^i \eta_i - \Psi(\theta).
\end{align}
The positive definite matrices $g_{ij}(\theta)$ and $g^{ij}(\eta)$ are obtained from the Hessian matrices
of the convex function $\Psi(\theta)$ and $\Psi^\star(\eta)$ as 
\begin{align} 
g_{ij} ( \theta) = 
\frac{\partial \eta_i}{\partial \theta^j} = 
\frac{\partial^2 \Psi( \theta)}{\partial \theta^i \partial \theta^j}, \quad
g^{ij} ( \eta) = 
\frac{\partial \theta^i}{\partial \eta_j} = 
\frac{\partial^2 \Psi^\star(\eta)}{\partial \eta^i \partial \eta^j}
\label{g} 
\end{align}
respectively.
These matrices satisfy
\begin{align}
g^{ij} (\eta) \,
g_{jk} (\theta) = 
\delta^i_k
\label{orthogonality}
\end{align}
where $\delta^i_k$ denotes Kronecker's delta.
The $\theta$- and $\eta$-coordinate systems are dual affine coordinates. 
Since connections are not tensors, there exists a coordinate system in which the connection becomes zero and such
a coordinate system is called \textit{affine coordinate}.

The $\alpha$-connection $\nabla^{(\alpha)}$ \cite{Amari}, which is a one-parameter family $ \left\{ \nabla^{\alpha} \right\}_{\alpha \in \mathbb{R}}$ of connections, and its dual $\nabla^{\star (\alpha)}$ are defined by their coefficients
\begin{align}
   \Christo^{(\alpha)}{}_{ijk}(\theta) := \frac{(1 - \alpha)}{2} C_{ijk}(\theta), \qquad
   \Christo^{\star (\alpha) \; ijk} (\eta) := \frac{(1 + \alpha)}{2} C^{ijk}(\eta)
   \label{aGamma}
\end{align}
respectively.
Here $C_{ijk}(\theta)$ and $C^{ijk}(\eta)$ are the total symmetric cubic tensors
\begin{align}
   C_{ijk}(\theta)  := \frac{\partial^3 \Psi(\theta)}{\partial \theta^i \partial \theta^j \partial \theta^k}, \qquad
   C^{ijk}(\eta)  := \frac{\partial^3 \Psi^\star(\eta)}{\partial \eta^i \partial \eta^j \partial \eta^k}
\end{align}
which are called \textit{Amari-Chentsov tensors}.
Note that when the parameter $\alpha\!=\!0$, both coefficients in \eqref{aGamma} reduce to the Christoffel symbol of the first kind $\Gamma_{ijk}$,
i.e., $\nabla^{(0)}$ is Levi-Civita connection.
Among the $\alpha$-connections, $\alpha = \pm 1$ play a central role \cite{Amari}. The coefficients of $\nabla^{(1)}$ and $\nabla^{\star (1)}$ in the $\theta$-coordinate system 
and those of $\nabla^{\star (-1)}$ and  $\nabla^{(-1)}$ in the $\eta$-coordinate system are as follows
\begin{align}
  \Christo^{(1)}{}_{ijk}(\theta) &= 0, \qquad \Christo^{(-1)}{}_{ijk}(\theta) = C_{ijk}(\theta) \\
  \Christo^{\star (1) \; ijk}(\eta) &= C^{ijk}(\eta), \qquad 
  \Christo^{\star (-1) \; ijk}(\eta) = 0. \nonumber
\end{align}
$\nabla^{(1)}$ is called \textit{exponential connection} and $\nabla^{(-1)}$ is called \textit{mixture connection}.
For more details for the basic of IG,
please refer to Amari's book \cite{Amari}. 
\smallskip 

\subsection{Gradient-Flow Equations}
The gradient-flow equations \cite{N94,FA95,BN16} in IG are brefly explained here.
The gradient-flow equations with respect to a convex $\Psi(\theta)$ function are 
given by
\begin{align} 
\frac{\rd \theta^i}{\rd t} = 
g^{ij} ( \theta) \,
\frac{\partial \Psi(\theta)}{\partial \theta^j}
\label{theta-gradEq}
\end{align}
in the $\theta$-coordinate system.
By using the properties \eqref{theta-eta} and \eqref{g}, the left-hand side (LHS) of \eqref{theta-gradEq} is rewritten by
\begin{align}
\frac{\rd \theta^i}{\rd t} &= 
\frac{\partial \theta^i}{\partial \eta_j} 
\frac{\rd \eta_i}{\rd t} = 
g^{ij} ( \theta) 
\frac{\rd \eta_i}{\rd t}
\end{align}
and the right-hand side (RHS) is
\begin{align} 
g^{ij} ( \theta) 
\frac{\partial \Psi(\theta)}{\partial \theta^j} = g^{ij} ( \theta) \, \eta_j.
\end{align}
Consequently, the gradient-flow equations \eqref{theta-gradEq} in the $\theta$-coordinate system are equivalent
to the linear differential equations
\begin{align}
\frac{\rd \eta_i}{\rd t} = 
\eta_i
\end{align}
in the $\eta$-coordinate system. This linearization is one of the merits due to the dually-flat structures \cite{Amari} in IG.

The other set of gradient-flow equations are given by
\begin{align} 
\frac{\rd \eta_i}{\rd t} = 
- g_{ij} ( \eta) 
\frac{\partial \Psi^\star(\eta)}{\partial \eta_j}
\label{eta-gradEq}
\end{align}
in the $\eta$-coordinate system.
Similarly, they are equivalent to the linear differential equations
\begin{align}
\frac{\rd \theta^i}{\rd t} = 
- \theta^i
\end{align}
in the $\theta$-coordinate system.
It is worth emphasizing that the two sets of the differential equations \eqref{theta-gradEq} and \eqref{eta-gradEq}
describe different processes in general \cite{WSM22, CW22}.

For later usage, we derive here the pre-geodesic (or non-affinely parametrized geodesic) equations for
the gradient-flow equations \eqref{theta-gradEq}. Taking the derivative of both sides of \eqref{theta-gradEq} with respect to $t$
we have
\begin{align}
 \label{non-affineEq}
 \frac{\rd^2 \theta^i}{\rd t^2} &=  \frac{\rd}{\rd t} \left( g^{i \ell}(\theta) \eta_{\ell} \right) = \frac{\partial g^{i \ell} (\theta) }{\partial \theta^k} \frac{\rd \theta^k}{\rd t} \eta_{\ell}
 + g^{i \ell} (\theta) \frac{\partial \eta_{\ell}}{\partial \theta^k}  \frac{\rd \theta^k}{\rd t}  \\
  &= \frac{\partial g^{i \ell} (\theta) }{\partial \theta^k} \frac{\rd \theta^k}{\rd t}  g_{\ell j}(\theta) \frac{\rd \theta^{j}}{\rd t}
 + \delta^i_k \frac{\rd \theta^k}{\rd t} = - g^{i \ell}(\theta) \frac{\partial g_{\ell j} (\theta) }{\partial \theta^k} \frac{\rd \theta^j}{\rd t}  \frac{\rd \theta^k}{\rd t}
 +  \frac{\rd \theta^i}{\rd t}  \nonumber \\
 \Leftrightarrow& \qquad 
  \frac{\rd^2 \theta^i}{d t^2}+g^{i\ell} (\theta) \frac{\partial g_{\ell j}(\theta) }{\partial \theta^k} \frac{\rd \theta^j}{\rd t} \frac{\rd \theta^k}{\rd t}
 = \frac{\rd \theta^i}{\rd t} \nonumber 
\end{align}
where the relation \eqref{g} and $(\partial g^{i \ell}(\theta) / \partial \theta^k) g_{\ell j}(\theta) = -g^{i \ell}(\theta) \,  \partial g_{\ell j}(\theta) / \partial \theta^k$ are used.
The last equations in \eqref{non-affineEq} are the  pre-geodesic equations
\begin{align}
  \frac{\rd^2 \theta^i}{\rd t^2} + \Christo^{(-1) \; i}{}_{j k}(\theta) \frac{\rd \theta^j}{\rd t} \frac{\rd \theta^k}{\rd t}
 = \frac{\rd \theta^i}{\rd t}
\end{align}
in the $\theta$-coordinate system. Here
\begin{align}
  \Christo^{(-1) \; i}{}_{j k}(\theta)  = g^{i\ell} (\theta) \Christo^{(-1)}{}_{\ell j k}(\theta) = g^{i\ell} (\theta) \frac{\partial g_{\ell j}(\theta) }{\partial \theta^k}
\end{align}
are the coefficients of the mixture ($\alpha \! = \! -1$) connection $\nabla^{(-1)}$.

\section{Weyl Integrable Geometry}
Here, some basics of Weyl integrable geometry \cite{PS14,RFP} are reviewed.
Weyl integrable geometry $(\mathcal{M}, g, \omega_k)$ is a generalization of Riemann geometry. In Riemann geometry, the metricity condition: $\nabla g = 0$ is satisfied, i.e., the covariant derivative $\nabla$ of a metric tensor $g$ equals zero,  In contrast, in Weyl integrable geometry we assume that
\begin{align}
   \Wnabla_{k} g_{i j} = \omega_{k} \, g_{i j}, \qquad  \Wnabla_{k} g^{i j} = -\omega_{k} \, g^{i j}
   \label{non_metricity}
\end{align}
which are called \textit{Weyl's non-metricity conditions} \cite{PS14}. 
Here the Weyl covector $\omega_{k} = \partial_k \omega$ denotes the $k$-th component of a one-form field $\omega$ on the manifold $\mathcal{M}$.
Using the relation of $\Wnabla_k (g_{i\ell} \; g^{\ell j}) = \Wnabla_k \delta_i^j = 0$, the first and second relations
in \eqref{non_metricity} are mutually related. As a result, the first relation is obtained from the second relation and vice versa.
The relations in \eqref{non_metricity} can be solved for the connection $\Wnabla$, which is called \textit{Weyl's connection}. 
The coefficients $\WChrist{}^{k}{}_{i j}$ of Weyl's connection $\Wnabla$ are derived from \eqref{non_metricity} as \cite{PS14} 
\begin{align}
 \WChrist{}^{k}{}_{i j} = \Christo^{k}{}_{i j} - \frac{1}{2} \left( \omega_{i} \, \delta^{k}_{j}+ \omega_{j} \, \delta^{k}_{i} 
   -  \omega^{k} \,  g_{i j} \right)
   \label{Wconnection}
 \end{align}
where $\Christo^{k}{}_{i j}$ denotes the coefficients of Levi-Civita connection with respect to a metric $g$ and $\omega^k = g^{k \ell} \omega_{\ell}$.
In the case of $\omega_k=0$, we recover Riemann geometry.

Next, for any smooth curve $C=C(\tau)$ and any pair of two parallel vector fields $V$ and $U$ along $C$, we have
\begin{align}
  \frac{\rd}{\rd \tau} g(V, U) = \omega \left( \frac{\rd}{\rd \tau }\right) \, g(V, U)
  \label{dg}
\end{align}
where $ \rd / \rd \tau$ denotes the vector tangent to $C$ and $\omega ( \rd /  \rd \tau)$ indicates the application of the one-form $\omega$ on  $ \rd / \rd \tau$.
In a coordinate basis
\begin{align}
  \frac{\rd} {\rd \tau} = \frac{\rd x^{j}}{\rd \tau} \frac{\partial }{\partial x^{j}},  \qquad \omega = \omega_{k} \rd x^{k}, \qquad
 V = V^{i} \frac{\partial}{\partial x^{i}}, \qquad U = U^{j} \frac{\partial}{\partial x^{j}}
 \end{align}
the relation \eqref{dg} becomes
\begin{align}
  \frac{\rd} {\rd \tau}  \left( g_{i j} V^{i} U^{j} \right) =  \omega_{k} \frac{\rd x^{k}}{\rd \tau} \, g_{i j} V^{i} U^{j}.
\end{align}
For the given Weyl covector $\omega_k$ satisfying Weyl's non-metricity conditions \eqref{non_metricity}, the most general expression of the pre-geodesic equations (Eq. 17 in \cite{PS14}) are given by
\begin{align}
  \frac{\rd x^j}{\rd \tau} \Wnabla_j \frac{\rd x^i}{\rd \tau} = \frac{1}{2 u^2} \left( \frac{\rd u^2}{\rd \tau} - u^2 \omega_j \frac{\rd x^j}{\rd \tau}  \right) \frac{\rd x^i}{\rd \tau}
  \label{WgeodesicEq}
\end{align}
where the tangent vector $u^{j} := \rd x^{j} / \rd \tau$ and
\begin{align}
  u^2 := g_{ij} \frac{\rd x^i}{\rd \tau} \frac{\rd x^j}{\rd \tau}.
\end{align}

Next, we see that the relation \eqref{non_metricity} is invariant under the following transformations with an arbitrary one-form $\Lambda$
\begin{subequations}
\label{gaugeT}
\begin{empheq}[left=\empheqlbrace]{align}
    g_{ij} \to \bar{g}_{ij} &= \exp(\Lambda) \, g_{ij} \\
    \omega_{k} \to \bar{\omega}_k &= \omega_k + \partial_k \Lambda
 \end{empheq}
\end{subequations}
which are called \textit{Weyl's gauge transformations}.
One can readily check that Weyl's connection \eqref{Wconnection} is also invariant under the gauge transformations \cite{PS14, RFP}.

For the given Weyl covector $\omega_{k}$ and tangent vector $v^{\nu} := \rd x^{\nu} / \rd \sigma$, the parameter $\sigma$ is called a \textit{proper time} \cite{ADR18} in Weyl geometry if it satisfies
 \begin{align}
    \frac{\rd v^2}{\rd \sigma} = v^2 \omega_{k} \frac{\rd x^{k}}{\rd \sigma}
    \label{Proptime}
 \end{align}
 where $v^2 := g_{i j} v^{i} v^{j}$.
 This condition \eqref{Proptime} is equivalent to
 \begin{subequations}
 \begin{empheq}[left=\empheqlbrace]{align}
    \rd v^2 &= v^2 \omega_{k} \rd x^{k} \\
    v^{k} &= \frac{\rd x^{k}}{\rd \sigma} \cdot
 \end{empheq}
 \end{subequations}
 Note that the tangent vector $v^{k}$ depends on the parameter $\sigma$.
 If the parameter $\tau$ in \eqref{WgeodesicEq} is a proper time, then the RHS of  \eqref{WgeodesicEq} becomes zero
 and \eqref{WgeodesicEq} is the geodesic equation. The parameter $\tau$ in this case is also called \textit{affine parameter}.

\section{Weyl Approach to the Gradient-Flow Equations}

Now, we come to the main section of this paper and the gradient-flow equations are reconsidered here from the perspective of Weyl integrable geometry.
From the gradient-flow equations \eqref{theta-gradEq} and using the second relation in \eqref{theta-eta}, we have
\begin{align}
  \frac{\rd \theta^i}{\rd t} = g^{ij}(\theta) \, \eta_j
  \label{constitutive}
\end{align}
which can be regarded as the relation between the velocity vector (or tangent vector) $\rd \theta^i / \rd t \in T\mathcal{M}$ (the tangent space) on a manifold $\mathcal{M}$ (the $\theta$-coordinate system)
and the momentum covector $\eta_i \in T^\star \mathcal{M}$ (the cotangent space). By using \eqref{constitutive} we obtain \cite{WSM21} that
\begin{align}
  g_{ij}(\theta) \frac{\rd \theta^i}{\rd t} \frac{\rd \theta^j}{\rd t} = g^{ij}(\theta) \eta_i(\theta) \eta_j(\theta) =: \eta^2(\theta)
  \label{eta2}
\end{align}
where $\eta^2(\theta)$ denotes the square of the length $\vert \eta(\theta) \vert$ as a function of the vector $\theta$.
In other words, the length $\vert \rd \theta / \rd t \vert$ of the velocity vector $\rd \theta / \rd t $ and the length $\vert \eta \vert$ of the momentum covector $\eta$ are same but their directions are different in general.

We next introduce the conformal metric $\tilde{g}$ which is related to the metric $g$ by the conformal transformation
\begin{align}
  \tilde{g}_{ij}(\theta) = \exp(\Lambda) g_{ij}(\theta) = \eta^2(\theta) \, g_{ij}(\theta)
\end{align}
where $\Lambda =  \ln \eta^2(\theta)$ is the associated Weyl one-form.
Then for the Riemann geometry $(\mathcal{\tilde{M}}, \tilde{g} )$,
the conformal metric $\tilde{g}$, of course, satisfies the metricity condition:
 \begin{align}
  \tilde{ \nabla}_k \, \tilde{g}_{ij}(\theta) = 0
 \end{align}
where $\tilde{ \nabla}$ is Levi-Civita connection with respect to the metric $\tilde{g}$.
Then, it follows that
\begin{align}
  0 &=\tilde{\nabla}_k \, \tilde{g}_{ij}(\theta) = \frac{\partial \Lambda}{\partial \theta^k} \, \exp(\Lambda) \, g_{ij}(\theta) + \exp(\Lambda) \tilde{\nabla}_k \, g_{ij} (\theta)
 \\
  &= \exp(\Lambda) \left( \frac{\partial \ln \eta^2(\theta)}{\partial \theta^k} \; g_{ij}(\theta) +  \tilde{\nabla}_k g_{ij}(\theta)  \right).  \nonumber 
\end{align}
Consequently, we have
\begin{align}
   \tilde{\nabla}_k g_{ij}(\theta) = - \frac{\partial \ln \eta^2(\theta)}{\partial \theta^k} \, g_{ij}(\theta) = \omega_k \, g_{ij}(\theta)
   \label{WnonMet}
 \end{align}
and the associated Weyl covector is identified with  $\omega_k = - \partial \ln \eta^2(\theta) /  \partial \theta^k$.  In this way, we can regard the $\theta$-coordinate space as the Weyl integrable geometry $(\mathcal{M}, g, \omega_k =  - \partial \ln \eta^2(\theta) /  \partial \theta^k) $, which is the main result of this work. Recall that
the connection $\tilde{\nabla}$ is Weyl's gauge invariant under the Weyl gauge transformations \eqref{gaugeT}.

For the tangent vector $u^i = \rd \theta^i / \rd t$, we have
\begin{align}
  u^2 = g_{ij} \frac{\rd \theta^i}{\rd t}  \frac{\rd \theta^j}{\rd t} = \left( \frac{\rd s}{\rd t} \right)^2 =  \eta^2(\theta)
  \label{u2}
\end{align}
and
\begin{align}
  \frac{\rd u^2}{\rd t} = \frac{\partial \eta^2(\theta)}{\partial \theta^k}  \frac{\rd \theta^k}{\rd t} 
  = u^2 \left(  \frac{\partial \ln \eta^2(\theta)}{\partial \theta^k } \right) \frac{\rd \theta^k}{\rd t} \cdot
  \label{thisu2}
\end{align}
Note that since the associated Weyl covector is $\omega_k = - \partial \ln \eta^2(\theta) /  \partial \theta^k$, by comparing this relation \eqref{thisu2} to \eqref{Proptime}  we see that the parameter $t$ is
not a proper time (or not an affine parameter).

Next, we consider the pre-geodesic equations \eqref{WgeodesicEq} in this Weyl Integrable geometry $(\mathcal{M}, g, \omega_k =  -\partial \ln \eta^2(\theta) /  \partial \theta^k )$ of the $\theta$-coordinate system.
\begin{align}
  &\frac{\rd \theta^j}{\rd t} \tilde{ \nabla}_j \frac{\rd \theta^i}{\rd t} = 
  \frac{1}{2 u^2} \left( \frac{\rd u^2}{\rd t} - u^2 \omega_j \frac{\rd \theta^j}{\rd t}  \right) \frac{\rd \theta^i}{\rd t} \nonumber \\
&=  \left( \frac{1}{ \eta^2(\theta)}  \frac{\partial \eta^2(\theta) }{\partial \theta^{\ell}  }  \right) \frac{\rd \theta^{\ell}}{\rd t}  \frac{\rd \theta^i}{\rd t}
 = \frac{1}{\eta^2({\theta})} \frac{\rd \eta^2(\theta)}{\rd t} \frac{\rd \theta^i}{\rd t} \cdot
  \label{this}
\end{align}
The LHS is
\begin{align}
  \frac{\rd \theta^j}{\rd t} \tilde{ \nabla}_j \frac{\rd \theta^i}{\rd t} = \frac{\rd \theta^j}{\rd t} \left( \frac{\partial }{\partial \theta^j} \frac{\rd \theta^i}{\rd t}  + \tilde{\Gamma}^i{}_{j k} \frac{\rd \theta^k}{\rd t} \right) 
  = \frac{\rd^2 \theta^i}{\rd t^2}  + \tilde{\Gamma}^i{}_{j k} \frac{\rd \theta^j}{\rd t} \frac{\rd \theta^k}{\rd t}
\end{align}
where
 \begin{align}
 \tilde{\Christo}^i{}_{j k} \; \frac{\rd \theta^j}{\rd t}  \frac{\rd \theta^k}{\rd t} 
 = \Christo^i{}_{j k} \;  \frac{\rd \theta^j}{\rd t} \frac{\rd \theta^k}{\rd t} 
  + \frac{1}{\eta^2(\theta)} \frac{\rd \eta^2(\theta)}{\rd t} \frac{\rd \theta^i}{\rd t} - \frac{1}{2} g^{i \ell}(\theta) \frac{ \partial \eta^2(\theta)}{\partial \theta^{\ell}}  \cdot
   \end{align}
Then, the pre-geodesic equations \eqref{this} are rewritten as
  \begin{align}
 \frac{\rd^2 \theta^i}{\rd t^2} +  \Christo^i{}_{j k} \;  \frac{\rd \theta^j}{\rd t} \frac{\rd \theta^k}{\rd t} 
  = \frac{1}{2} g^{i \ell}(\theta) \frac{ \partial \eta^2(\theta)}{\partial \theta^{\ell}}  \cdot
  \label{geo}
   \end{align}
 From the relation \eqref{u2} and the gradient-flow equations \eqref{theta-gradEq}
 \begin{align}
   \frac{\rd \theta^i}{\rd t} = g^{ij}(\theta) \, \eta_j
 \end{align}
 we obtain
 \begin{align}
 \frac{1}{2} \frac{\partial \eta^2(\theta)}{\partial \theta^i} = \eta_i - \frac{1}{2} \frac{ \partial g_{jk}(\theta) }{\partial \theta^i} \frac{\rd \theta^j}{\rd t} \frac{\rd \theta^k}{\rd t} \cdot
 \end{align}
 Then, the RHS of \eqref{geo} becomes
 \begin{align}
  g^{i \ell}(\theta) \left(  \eta_{\ell} - \frac{1}{2} \frac{\partial g_{jk}(\theta)}{\partial \theta^{\ell} }\frac{\rd \theta^j}{\rd t} \frac{\rd \theta^k}{\rd t} \right) 
  = \frac{\rd \theta^i}{\rd t} - \frac{g^{i \ell}(\theta)}{2} \frac{ \partial g_{jk}(\theta)}{\partial \theta^{\ell}} \frac{\rd \theta^j}{\rd t} \frac{\rd \theta^k}{\rd t} \cdot
 \end{align}
 Since the coefficients of Levi-Civita connection $\nabla^{(0)}$ with respect to the metric $g$ are
\begin{align}
  \Christo^i{}_{j k} = \frac{g^{i \ell}(\theta) }{2} \left( \frac{\partial g_{\ell k}(\theta) }{\partial \theta^j} + \frac{\partial g_{j \ell} (\theta) }{\partial \theta^k} - \frac{\partial g_{jk}(\theta) }{\partial \theta^{\ell} }\right)
\end{align}
the pre-geodesic equations \eqref{geo} are equivalent to
 \begin{align}
  \frac{\rd^2 \theta^i}{\rd t^2} + \Christo^{ \textrm{(-1)}}{}^i{}_{jk} \;  \frac{\rd \theta^j}{\rd t} \frac{\rd \theta^k}{\rd t} = \frac{\rd \theta^i}{\rd t}
\end{align}
where $ \Christo^{\textrm{(-1)}}{}^i{}_{jk}$ are the coefficients of the mixture ($\alpha \!=\! -1$) connection $\nabla^{(-1)}$, i.e.,
 \begin{align}
   \Christo^{\textrm{(-1)}} {}^i{}_{jk} =  \frac{g^{i \ell}(\theta)  }{2} \left( \frac{\partial g_{\ell k}(\theta) }{\partial \theta^j} + \frac{\partial g_{j \ell}(\theta) }{\partial \theta^k}  \right)
   = g^{i \ell}(\theta)  C_{\ell j k}(\theta).
\end{align}
In this way we have shown that the pre-geodesic equations associated with the gradient-flow equations are
the pre-geodesic equations in Weyl integrable geometry $(\mathcal{M}, g, \omega_k =  -\partial \ln \eta^2(\theta) /  \partial \theta^k )$.
Note that the associated Weyl one form $\ln \eta^2(\theta)$ plays a central role and it is related to the $\theta$-potential function $\Psi(\theta)$ as follows.
\begin{align}
  \frac{\rd \Psi(\theta)}{\rd t} = \frac{\partial \Psi(\theta)}{\partial \theta^i} \frac{\rd \theta^i}{\rd t} = \eta_i \frac{\rd \theta^i}{\rd t} = g^{ij}(\theta) \eta_i \eta_j
  =  \eta^2(\theta)
  \label{dPsi_dt}
\end{align}
where the relations \eqref{theta-eta}, \eqref{constitutive} and \eqref{eta2} are used. We emphasize that the scalar field $ \eta^2(\theta)$ characterizes
the rate of the $\theta$-potential. 

Needless to say that the similar argument applies to the other set of the gradient-flow equations \eqref{eta-gradEq} in the $\eta$-coordinate system.
For example, the corresponding relation of \eqref{dPsi_dt} are
\begin{align}
 - \frac{\rd \Psi^{\star}(\eta)}{\rd t} =  -\frac{\partial \Psi^{\star}(\eta)}{\partial \eta_i} \frac{\rd \eta_i}{\rd t} = -\theta^i \frac{\rd \eta_i}{\rd t} 
  = g_{ij}(\eta) \theta^i \theta^j =  \theta^2(\eta).
\end{align}
Since $-\Psi^{\star}(\eta)$ is the entropy function, the LHS is the entropic rate in the entropic gradient-flows. Then it is found that this entropy rate 
$ - \rd \Psi^{\star}(\eta) / \rd t$ is characterized by the scalar field $\theta^2(\eta)$, which is an important ingredient in the Weyl integral geometry.

\section{Conclusions}
We have reconsidered the gradient-flow equations in IG from the perspective of the Weyl integrable geometry.
The gradient-flow equations are related to the pre-geodesic equations in the Weyl integrable geometry. 
In conventional way \cite{Amari} of IG, the natural coordinate ($\theta$- and $\eta$-) spaces are characterized with $\alpha$-connections $\nabla^{(\alpha)}$, which provide the parallel translation rule in these spaces. In addition, unlike Riemann geometry, the Hessian metics $g$ in \eqref{g} are used to determine the orthogonality \eqref{orthogonality} only.
The $\theta$- and $\eta$-coordinate systems are regarded as the different coordinates on the same manifold.
This is the perspective of  modern affine geometry, in which the physical meaning of $\alpha$-connections seems to be not clear unfortunately.

In contrast,  in the perspective of the Weyl integrable geometry in this work,  the $\theta$- and $\eta$-coordinate systems are regarded as
the different manifolds. For example,
$g_{ij}(\theta)$ in \eqref{g} are the components of the  Riemann metric on the $\theta$-coordinate space $\mathcal{M}$, and the Wely connection $\tilde{\nabla}$ in \eqref{WnonMet} provides the parallel translation rule. The Wely covector $\omega_k =  - \partial \ln \eta^2(\theta) /  \partial \theta^k$ is characterized by the scalar field $\eta^2(\theta)$ in \eqref{eta2}.
In this way, the $\theta$-coordinate space can be characterized with the metric $g_{ij}(\theta)$,  the Weyl connection $\tilde{\nabla}$, and the Weyl covector $\omega_k$, i.e., the Weyl integrable geometry $(\mathcal{M}, g, \omega_k =  - \partial \ln \eta^2(\theta) /  \partial \theta^k) $.

\section*{Acknowledgements}
The author is partially supported by Japan Society for the Promotion
of Science (JSPS) 
Grants-in-Aid for Scientific Research (KAKENHI)
Grant Number 22K03431. 

\appendix

%

Region of Electrical and Electronic Systems Engineering, \\
Ibaraki University\\
316-8511 Hitachi,\\
Ibaraki, JAPAN \\
\textit{E-mail address}: \href{mailto:tatsuaki.wada.to@vc.ibaraki.ac.jp}{\nolinkurl{tatsuaki.wada.to@vc.ibaraki.ac.jp}}

\label{last}
\end{document}